\documentclass[a4paper, 12pt]{article}
\usepackage{amsmath,bbm,amssymb}
\usepackage{graphicx}
\usepackage{makeidx}

\author{Clovis Jacinto de Matos\footnote{ESA-HQ, European Space Agency, 8-10 rue Mario Nikis,
75015 Paris, France, e-mail: Clovis.de.Matos@esa.int}}

\title{Testing Loop Quantum Gravity and Electromagnetic Dark Energy in Superconductors}

\begin{document}

\maketitle \begin{abstract}In 1989 Cabrera and Tate reported an
anomalous excess of mass of the Cooper pairs in rotating thin
Niobium rings. So far, this experimental result never received a
proper theoretical explanation in the context of superconductor's
physics. In the present work we argue that what Cabrera and Tate
interpreted as an anomalous excess of mass can also be associated
with a deviation from the classical gravitomagnetic Larmor theorem
due to the presence of dark energy in the superconductor, as well
as with the discrete structure of the area of the superconducting
Niobium ring as predicted by Loop Quantum Gravity. From Cabrera
and Tate measurements we deduce that the quantization of spacetime
in superconducting circular rings occurs at the Planck-Einstein
scale $l_{PE} = (\hbar G/c^3 \Lambda)^{1/4}\sim 3.77\times 10
^{-5} m$, instead of the Planck scale $l_{P} =(\hbar G /
c^3)^{1/2}=1.61 \times 10 ^{-35} m$, with an Immirzi parameter
which depends on the specific critical temperature of the
superconducting material and on the area of the ring. The
stephan-Boltzmann law for quantized areas delimited by
superconducting rings is predicted, and an experimental concept
based on the electromagnetic black-body radiation emitted by this
surfaces, is proposed to test loop quantum gravity and
electromagnetic dark energy in superconductors.
\end{abstract}

\section{Introduction} The physics of SuperConductors (SC) could be
relevant for Loop Quantum Gravity (LQG) with respect to
investigating:
\begin{enumerate}
\item\label{i1} The scale at which spacetime acquires a discrete
structure.

\item \label{i2} The possibility of a spontaneous breaking of the
Principle of General Covariance (PGC) in SCs.

\item \label{i3} The thermal time hypothesis and the physical
nature of time in SCs.

\end{enumerate}
\section{Anomalous Cooper pair mass excess}
\begin{figure}[h!]
\begin{center}
\includegraphics[scale=0.5]{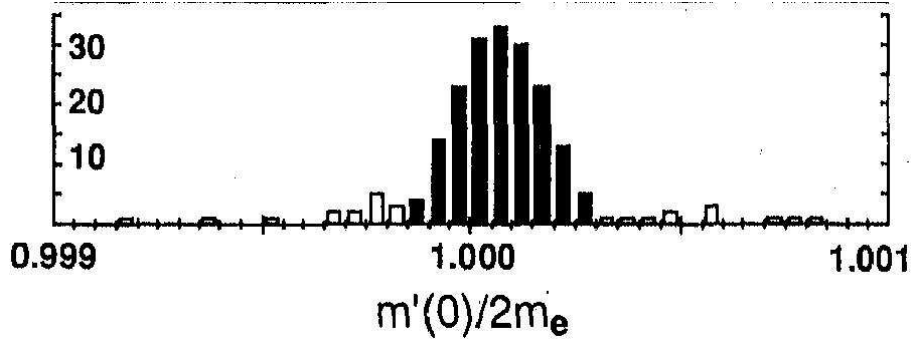}
\caption{\label{Cabrera1} Bar graph showing the distribution of
all data points of the Cooper pair relative mass excess,
$m^*/2m_e$, in Cabrera and Tate experiment, for Nb ring's rotation
frequency intervals of 2 Hz. We see in the present case that the
average value of the relative mass excess is
$m^*/2m_e=1.000075(25)$ \cite{Tate01}}
\end{center}
\end{figure}

In 1989 Cabrera and Tate \cite{Tate01, Tate02}, through the
measurement of the magnetic trapped flux originated by the London
moment, reported an anomalous Cooper pair mass excess in thin
rotating Niobium superconductive rings:
\begin{equation}
\Delta m=m^*-m=94.147240(21)eV\label{e0}
\end{equation}
Here $m^*=1.000084(21)\times 2m_e=1.023426(21)MeV$ is the
experimentally measured Cooper pair mass (with an accuracy of 21
ppm), see Figure \ref{Cabrera1} displaying a typical data set, and
$m=0.999992\times2m_e=1.002331 MeV$ is the theoretically expected
Cooper pair mass including relativistic corrections. We can also
express the relative excess of mass as:
\begin{equation}
\frac{\Delta m}{m}=9,2\times10^{-5}\label{e1}
\end{equation}

The above Cooper pair mass excess (or, equivalently, the slightly
larger than expected measured magnetic field) has not been
explained until now in the context of superconductor's physics.

The principle of Cabrera and Tate experiment, is to measure the
magnetic flux originated from the London moment which is trapped
in a thin Niobium ring.

Integrating the current density of Cooper pairs around a closed
path including the effect of a rotating frame:
\begin{equation}
\frac{m^*}{e^2 n_s} \oint_{\Gamma} \vec j \cdot \vec{dl}=n
\frac{h}{2e} - \int_{S_\Gamma} \vec B \cdot \vec{dS}-
\frac{2m^*}{e} \vec{\omega}\vec{S_\Gamma} \label{e2}
\end{equation}
Where $n_s$ is the Cooper pair number density, $S_\Gamma$ is the
area bounded by the closed curve, $\Gamma$, circulating inside the
superconductor, $\omega$ is the SC's angular velocity,
$B=-\frac{m}{e} 2 \omega$ is the London moment.

There exist an angular velocity $\omega_n$ for each $n$ such that
the flux $\int\vec{B}\cdot\vec{dS}$ and the electric current line
integral $\oint_{\Gamma} \vec j \cdot \vec{dl}$ are zero together,
see Figure \ref{Cabrera2}. This allows to define the flux null
spacing by
\begin{equation}
\Delta \omega \equiv \omega_n - \omega_{n-1}\label{e3}
\end{equation}
Subtracting eq.({\ref{e2}) for $n$ and $n-1$ we obtain:
\begin{equation}
\frac{h}{m^*}=S_\Gamma \Delta \omega\label{e4}
\end{equation}
Which is the key formula used by Cabrera and Tate to calculate the
Cooper pair's mass.

\begin{figure}[h!]
\begin{center}
\includegraphics[scale=0.5]{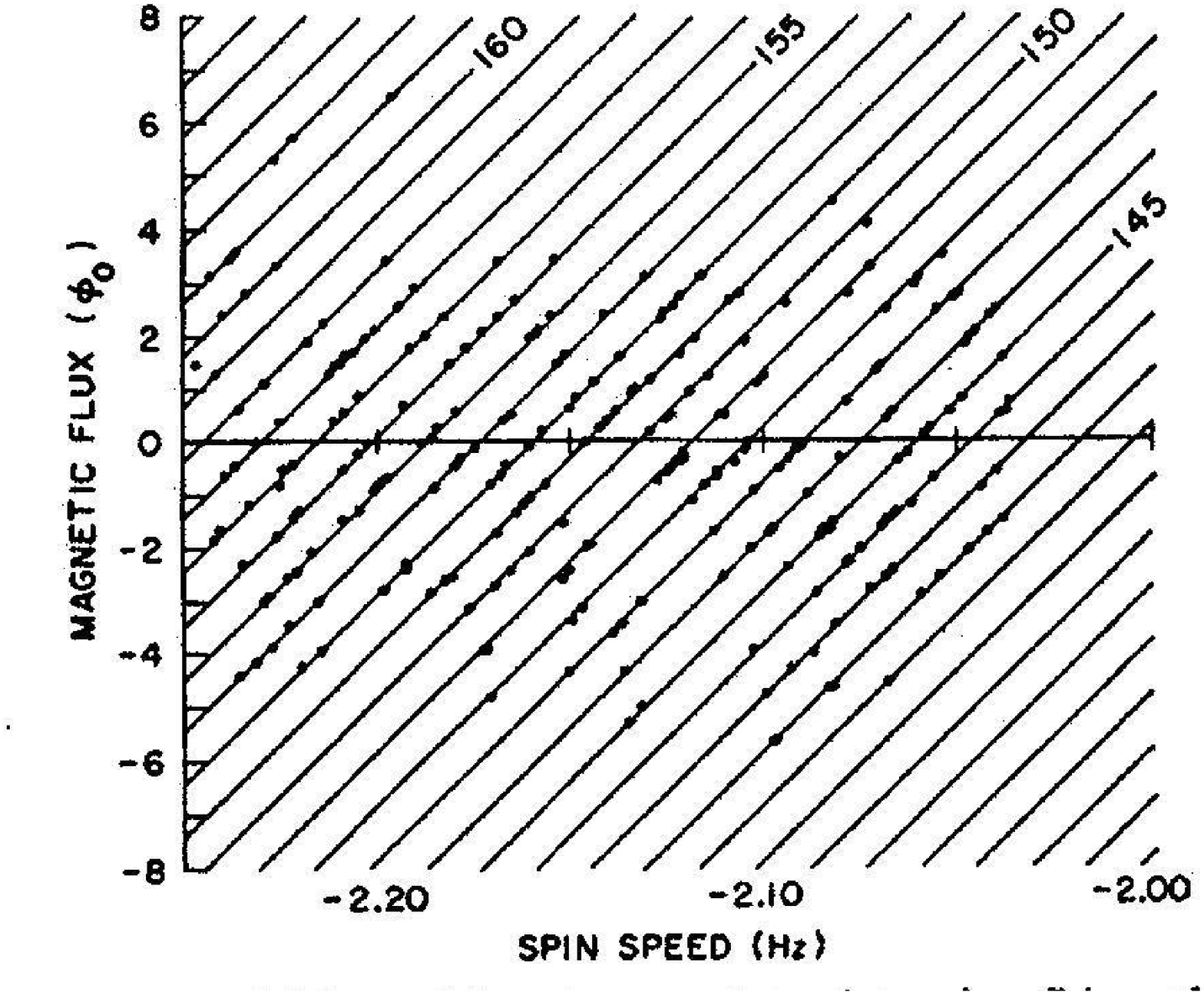}
\caption{\label{Cabrera2} Measured magnetic flux as a function of
angular velocity for different quantum states
$n=..160..155..150..145..$ of a rotating thin Niobium
superconduting ring in Cabrera and Tate experiment \cite{Tate02} }
\end{center}
\end{figure}

\section{Possible interpretations of Cabrera and Tate experiment}
What are the physical parameters which should vary to offer
alternatives to the interpretation of Cabrera and Tate's
experiment in terms of an anomalous excess of mass? From
eq.({\ref{e4}) we see that we have two possibilities in addition
to the Cooper pair mass parameter:
\begin{enumerate}
\item\label{i4} The area of the Niobium ring is different when the ring is in the normal state and
when it is in the superconducting state.
\item \label{i5} We must add a gravitomagnetic term to $\Delta \omega$, eq.(\ref{e3}), i.e., we must
add a GM term in the Cooper pair canonical momentum.
\end{enumerate}

\subsection{Fundamental quantum of area in superconductors and the
Immirzi parameter} Let us first investigate possibility \ref{i4}).
Either we observe a real excess of mass of Cooper pairs, $m^*$,
and the area of the ring in the superconducting state is the same
as in the normal state:
\begin{equation}
m^*=\frac{h}{S_\Gamma \Delta \omega}\label{e5}
\end{equation}
Where $S_\Gamma$ will be called the Euclidian area of the SCing
ring's hole. Or the Cooper pair's mass remains unchanged and the
area of the SCing ring is different from its area in the normal
state:
\begin{equation}
m=\frac{h}{S'_\Gamma \Delta \omega}\label{e6}
\end{equation}
Where $S'_\Gamma$ will be called the non-Euclidean area of the
SCing ring's hole. Subtracting eq.(\ref{e6}) from eq.(\ref{e5}) we
find:
\begin{equation}
\frac{\Delta m}{m}=\frac{S'_\Gamma -
S_\Gamma}{S_\Gamma}=\frac{\Delta S_\Gamma}{S_\Gamma}\label{e7}
\end{equation}
Inserting numerical values from Cabrera and Tate experiment in
eq.(\ref{e7}) we find the fundamental quantum of area:
\begin{equation}
\Delta S_\Gamma=1,86\times10^{-7} [m^2]\label{e8}
\end{equation}

In LQG the principal series of eigenvalues of the area, is labeled
by multiplets of half integers $j_i, i=1,...,n$ and is given by:
\begin{equation}
A=8 \pi \gamma\Big(\frac{\hbar G}{c^3}\Big)
\sum_i\sqrt{j_i(j_i+1)}\label{e9}
\end{equation}
where $\gamma$, called the immirzi parameter, is a free
dimensionless constant of the theory \cite{Rovelli}. In
eq.(\ref{e9}) we are assuming that the fundamental quanta of area
for a given fundamental value of $j_i$ is proportional to the
Planck area, $l_P^2=\Big(\frac{\hbar G}{c^3}\Big)$. Therefore we
are assuming that the fundamental scale for quantum gravity is the
Planck scale. However Beck, Mackey and CDM have argued that the
fundamental scale for quantum gravity in superconductors should be
the Planck-Einstein scale corresponding to the geometric mean
between the Planck scale, $l_P=\Big(\frac{\hbar
G}{c^3}\Big)^{1/2}$ which determines the highest possible energy
in the universe, and the Einstein scale, $l_E=\Lambda^{-1/2}$,
which is determined by the non-zero value of the Cosmological
Constant (CC) $\Lambda$ and fixes the lowest possible energy in
the universe.
\begin{equation}
\l_{PE}=\sqrt{l_P l_E}= \Big(\frac{\hbar G}{c^3
\Lambda}\Big)^{1/4}\label{e10}
\end{equation}
This would lead to a fundamental quantum of area in SCs that would
be proportional to the Planck-Einstein area:
\begin{equation}
A_{PE}=l_{PE}^2=\sqrt{A_P A_E}=\Big(\frac{\hbar G}{c^3
\Lambda}\Big)^{1/2}\label{e11}
\end{equation}
substituting eq.(\ref{e11}) in the LQG expression for the area,
eq.(\ref{e9}) we get:
\begin{equation}
A=8 \pi \gamma\Big(\frac{\hbar G}{c^3 \Lambda}\Big)^{1/2}
\sum_i\sqrt{j_i(j_i+1)}\label{e12}
\end{equation}
for the fundamental value $j=1/2$ we have the fundamental quantum
of area:
\begin{equation}
A_{1/2}=4\sqrt3 \pi \gamma \Big(\frac{\hbar G}{c^3
\Lambda}\Big)^{1/2} \label{e13}
\end{equation}
Assuming $A_{1/2}=\Delta S_\Gamma$ we can substitute
eq.(\ref{e13}) in eq.(\ref{e7}) and use the measurement of Cabrera
and Tate for the Copper pairs excess of mass, as well as the
Euclidian dimensions of their SCing Niobium ring
$S_\Gamma=2,02\times10^{-3} m^2$, to find the value of the Immirzi
parameter in the SC:
\begin{equation}
\gamma\simeq6\label{e14}
\end{equation}

\subsection{Electromagnetic dark energy in superconductors}
Let us now investigate possibility \ref{i5}) above. In this case
the Cooper pair canonical momentum, $\vec \pi$, should include a
GravitoMagnetic (GM) term.
\begin{equation}
\vec {\pi}= m\vec v + e \vec A + m \vec {A_g}\label{e15}
\end{equation}
Where $\vec A$ is the magnetic vector potential, $v$ is the Cooper
pair velocity, and $\vec {A_g}$ is the GM vector potential, whose
rotational gives the GM field:
\begin{equation}
\vec {B_g}=\nabla \times \vec{A_g}\label{e16}
\end{equation}

The Ginzburg-Landau eq.(\ref{e2}) should now read as:
\begin{equation}
\frac{m}{e^2 n_s} \oint_{\Gamma} \vec j \cdot \vec{dl}=n
\frac{h}{2e} - \int_{S_\Gamma} \vec B \cdot \vec{dS}- \frac{m}{e}
\int_{S_\Gamma} \vec {B_g} \cdot \vec{dS}- \frac{2m}{e}
\vec{\omega}\vec{S_\Gamma} \label{e17}
\end{equation}

Subtracting eq(\ref{e2}) from eq.(\ref{e17}) we get:
\begin{equation}
\vec{B_g}=2 \vec \omega \Big(\frac{m^* - m}{m}\Big)+\Big(\frac{m^*
- m}{m}\Big) \frac{1}{S_\Gamma e n_s}\oint \vec j \cdot \vec{dl}
\label{e18}
\end{equation}
In a SC that is thick compared with the London penetration depth,
the circular path $\Gamma$ can be chosen in the SC's bulk where
there is no current flowing, thus leading to a null current
integral in eq.(\ref{e18}), and:
\begin{equation}
\frac{\Delta m}{m}=\frac{B_g}{2 \omega}\label{e20}
\end{equation}
Eq.(\ref{e20}) is interpreted as indicating a deviation with
respect to the classical GM Larmor theorem \cite{Tajmar01} \cite
{Tajmar02}.

Beck and CDM \cite{Beck01}\cite{de Matos03}\cite{de Matos04} have
discussed the possibility of that the source of the gravitational
and gravitomagnetic fields in SCs is the density of
electromagnetic zero point energy, $\rho^*$, contained in the SC.
Although this requires a spontaneous breaking of the Principle of
Generale possibility Covariance (PGC) when the material crosses
its critical temperature, $T_c$. Meaning that below $T_c$ the
density of electromagnetic zero point energy has the same physical
nature as the vacuum energy, $\rho_{CC}$ associated with the CC:
\begin{equation}
\rho_{CC}= \frac{c^4 \Lambda}{8 \pi G}\label{e21}
\end{equation}
and above $T_c$, $\rho^*$ does not contribute anymore the
cosmological vacuum energy, $\rho_{CC}$. Since the currently
measured value of the cosmological constant,
$\Lambda=1,29\times10^{-59} [m^{-2}]$, accounts reasonably for the
density of dark energy observed in the universe, we can say that
our model of gravitationally active electromagnetic zero point
energy in SCs is an electromagnetic model of dark energy in
superconducting matter. Each SCs would host a different density of
electromagnetic dark energy proportional to the fourth power of
its critical temperature \cite{Beck01}-\cite{Beck06}:
\begin{equation}
\rho^*=\frac{\pi \ln^4(3)}{2} \frac{k^4}{(ch)^3} T_c^4\label{e22}
\end{equation}
CDM \cite{de Matos03} has shown that the Cooper pairs excess of
mass measured by Cabrera and Tate is proportional to the ratio of
electromagnetic dark energy contained in the superconductor and
the cosmological density of dark energy:
\begin{equation}
\frac{\Delta m}{m}=\frac{B_g}{2 \omega}=\frac {3}{2}
\frac{\rho^*}{\rho_{CC}}\label{e23}
\end{equation}
In the case of Niobium eq.({23}) is in excellent agreement with
the measured value eq.(\ref{e1}).

In the  following we will assume that:

\begin{enumerate}
\item\label{i6} Cooper pairs mass excess,

\item \label{i7} discrete areas at the Planck-Einstein scale,

\item \label{i8} non-classical inertia in superconductors,

\end{enumerate}
are different equivalent phenomenological interpretations of the
spontaneous breaking of the PGC in SCs, i.e, we assume:
\begin{equation}
\frac{\Delta m}{m}=\frac{\Delta
S_\Gamma}{S_\Gamma}=\frac{B_g}{2\omega}=\frac{3}{2}\frac{\rho^*}{\rho_{CC}}=\chi\label{e24}
\end{equation}

\section{Discrete spacetime and electromagnetic dark energy}
To consolidate further the physical concept of a discrete
structure of spacetime at the Planck-Einstein scale in SCs
\cite{Beck01}, let us mention briefly how one can deduce SC's
inertial properties as resulting from quantum fluctuations of the
SC's four volume
\subsection{Uncertainty relations and discrete spacetime}
The successful resolution of the inverse CC problem \cite{Beck01}
encourages us to start from the assumption that the spacetime
volume of a superconductor is made of Planck-Einstein cells,
$l_{PE}^4$, which will fluctuate:
\begin{equation}
\Delta V\sim\sqrt V l_{PE}^2\label{e25}
\end{equation}
Since the density of vacuum energy associated with the CC,
$\rho_{CC}$ is canonically conjugated with the universe
four-volume $V$.
\begin{equation}
\Delta \rho_{CC} \Delta V \sim \hbar c \label{e26}
\end{equation}
we use the electromagnetic dark energy density, eq.(\ref{e22}) and
the SC's four volume eq.(\ref{e25}) instead of respectively
$\rho_{CC}$ and $\Delta V$ in eq.(\ref{e26}). In this way we
obtain that the inertia in superconducting matter changes with
respect to its classical laws due to the fluctuations of the SC's
discrete spacetime volume:
\begin{equation}
\Delta \chi \sqrt{V} \sim \frac{2\pi ^2}{3} l^2_{PE}\label{e27}
\end{equation}
Substituting eq.(\ref{e24})in eq.(\ref{e27}) we find
\begin{equation}
\Delta S_\Gamma \sqrt{V}\sim n \frac{2 \pi^2} {3}
l_{PE}^4\label{e28}
\end{equation}
Where $n=S_\Gamma / l_{PE}^2$ is the number of Planck-Einstein
quanta of area making the Euclidean value of $S_\Gamma$. This
could indicate the the fundamental value of the quantum of area of
surface delimited by SCs could fluctuate as a consequence of a
four-volume fluctuations of the SC.
\subsection{Stephan-Boltzmann law for discrete spacetime surfaces
in superconducting rings} Let us now use the electromagnetic dark
energy model in SCs to find a thermodynamical law of the Immirzi
parameter.

From the equality of $A_{1/2}$ and $S_{\Gamma}$, $A_{1/2}=\Delta
S_\Gamma$, and substituting eq.(\ref{e13}) in eq.(\ref{e7}), and
using eq.(\ref{e24})(\ref{e22})and (\ref{e21}) we obtain:
\begin{equation}
\gamma=\frac{3 \ln^4(3)}{32 \sqrt{3} \pi^2} \Big(
\frac{T_c}{T_{PE}}\Big)^4 \frac{S_\Gamma}{l_{PE}^2} = n \frac{3
\ln^4(3)}{32 \sqrt{3} \pi^2} \Big(
\frac{T_c}{T_{PE}}\Big)^4\label{e29}
\end{equation}
Where as above  $n=S_\Gamma / l_{PE}^2$ is the number of
Planck-Einstein quanta of area making the Euclidean value of
$S_\Gamma$, $T_{PE} = \frac{1}{k} \Big(\frac {c^7 h^3
\Lambda}{G}\Big)=60.71 K$ is the Planck-Einstein temperature.
Using the Euclidean value of the area of the Niobium SCing ring
used by Cabrera and Tate, $S_\Gamma=2,02\times10^{-3} m^2$, in
eq.(\ref{e29}) we find indeed the same result than in
eq.(\ref{e14}), i.e, $\gamma \sim 6$. Taking the same surface for
the different superconducting materials listed in table 1 below we
can calculate the respective Immirzi coefficients.

\begin{center}
\begin{tabular}{|c|c|c|}
\hline Superconductive material & $T_c [K]$ & $\gamma$ \\
\hline $Al$ & $1,18$ & $0,0016$ \\
\hline $Sn$ & $3,72$ & $0,16$ \\
\hline $Pb$ & $7,2$ & $2,24$ \\
\hline $Nb$ & $9,25$ & $6,11$ \\
\hline $Nb_3 G_2$ & $23,2$ & $242,14$ \\
\hline $YBCO$ & $91,0$ & $57316,81$ \\ \hline
\end{tabular}
\end{center}
Table 1: Immirzi parameter $\gamma$ predicted by the model of
electromagnetic dark energy in superconductors.
\bigskip

How do these values stand with respect to the usual discussion of
the Immirzi parameter in the context of Black-hole thermodynamics
\cite{Rovelli}? Bekenstein suggested that the horizon of a
Black-hole of mass, $M$, irradiates like a Black-Body at the
temperature
\begin{equation}
T=\frac{\hbar c^3}{a 32 \pi k G M}\label{f1}
\end{equation}
Where $a$ is a constant to be ultimately determined
experimentally, although LQG calculations predict a value of $a$
in function of the Immirzi parameter, $\gamma$:
\begin{equation}
a\sim \frac{0.2375}{4 \gamma}\label{f2}
\end{equation}
For $\gamma=0,2375$ we obtain $a=1/4$, which leads to the Hawking
prediction for the Black hole temperature:
\begin{equation}
T=\frac{\hbar c^3}{8 \pi k G M}\label{f3}
\end{equation}
This is the current procedure to fix the value of $\gamma$ in LQG.

Substituting eq.(\ref{f2}) in eq.(\ref{f1}) we get:
\begin{equation}
M=\frac{\hbar c^3 \gamma}{1.9 \pi k G T}\label{f4}
\end{equation}
Using the values for $\gamma$ and $T=T_c$ listed in Table 1) we
can estimate what would be the mass of the black-hole which would
be needed to generate the quantum gravitational effects, at the
Planck scale, that we predict in superconducting rings, with area
$S_\Gamma=2,02\times10^{-3} m^2$, at the Planck-Einstein scale.
The result is listed in table 2

\begin{center}
\begin{tabular}{|c|c|c|c|}
\hline Superconductive material & $T_c [K]$ & $\gamma$ & $M_{Black-Hole} [Kg]$\\
\hline $Al$ & $1,18$ & $0,0016$ & $7,1\times 10^{20}$\\
\hline $Sn$ & $3,72$ & $0,16$ & $2,21\times 10^{22}$\\
\hline $Pb$ & $7,2$ & $2,24$ & $1,61\times 10^{23}$\\
\hline $Nb$ & $9,25$ & $6,11$ & $3,42\times 10^{23}$\\
\hline $Nb_3 G_2$ & $23,2$ & $242,14$ & $5,4\times 10^{24}$\\
\hline $YBCO$ & $91,0$ & $57316,81$ & $3,26\times 10^{26}$\\
\hline
\end{tabular}
\end{center}
Table 2: Equivalent black hole mass (Classical physical system at
the Planck scale) corresponding to the LQG Immirzi parameter
$\gamma$ predicted by the model of electromagnetic dark energy in
superconductors at the Planck-Einstein scale.
\bigskip

We see that the masses of the corresponding black-holes ranges
from $M_{Moon}/100$ until $4436 M_{Moon}$. This "boost" of quantum
gravitational effects in superconductors in the Earth laboratory,
which have masses much smaller than the Moon's mass, would be due
to the important ratio between the Planck-Einstein scale and the
Planck scale. The cause of this scale transformation could be a
spontaneous breaking of the PGC in superconductors.

Let us now investigate the possible physical phenomenology
associated with eq.(\ref{e29}). Defining the constant $\kappa
[m^{-2}K^{-4}]$:
\begin{equation}
\kappa=\Big(\frac{3 \ln^4(3)}{32 \sqrt{3} \pi^2} \frac{1}{T_{PE}^4
l_{PE}^2}\Big)^{-1} \label{e30}
\end{equation}
we can re-write eq.(\ref{e29}) in a form analog to the
Stephan-Boltzmann law.
\begin{equation}
\gamma=\kappa S_\Gamma T_c^4\label{e31}
\end{equation}
The total electromagnetic power $\wp$ irradiated by a black body
is given by the Stephan-Boltzmann law:
\begin{equation}
\wp=\sigma S T^4\label{e32}
\end{equation}
Where $\sigma=5,67\times 10^{-8} [J/s m^2 K^4]$ is the
Stephan-Boltzmann constant, $S$ is the irradiating area of the
black body, and T is its temperature.

Making equal eq.(\ref{e31}) and eq.(\ref{e32}), $S_\Gamma
T_c^4=ST^4$, we deduce.
\begin{equation}
\wp=\gamma \frac{\sigma}{\kappa}\label{e33}
\end{equation}
substituting eq.(\ref{e31}) in eq.(\ref{e33}) we find:
\begin{equation}
\wp=\sigma S_\Gamma T_c^4\label{e34}
\end{equation}
Which is the Stephan-Boltzmann law for the area delimited by the
SCing ring, i.e. the SCing ring's hole. At temperatures of the
order of $T_c\sim 2,73$ it would be difficult to distinguish the
thermal radiation coming from the ring's hole from the one coming
from the Cosmic Microwave Background (CMB). In the following
section an experimental concept to detect the Black-Body thermal
radiation coming from areas delimited by superconductors, is
proposed.

\section{Experimental concept to test LQG and electromagnetic dark
energy in superconductors} The experimental concept to test LQG
and dark energy according to the previous discussion, is to
measure the black-body radiation of a thin SCing ring located near
a parabolic reflector, and concentrate the electromagnetic
radiation at the reflector's foci, where it would be detected. The
goal would be to exhibit an anomalous excess of thermal energy
originated from the SCing ring (empty) hole according to
eq.(\ref{e34}). In order to achieve this measurement one needs, at
least, to subtract from the total amount of thermal energy
detected:
\begin{enumerate}
\item The Cosmic microwave background,

\item The black body radiation coming from the reflector.
\end{enumerate}
The reflector should not be in a superconducting state to avoid
creating additional black-body radiating geometrical surfaces.

By measuring the spectral composition of the Planckian radiation
emitted by the SCing ring's hole, one could test the predictions
of LQG about the non-existence of the Bekenstein-Mukhanov effect
on Hawking's thermal radiation emitted by black holes. This
spectral analysis could also contribute to a better understanding
of the spin-network underlying the surface.

\section{Conclusions}
A bi-dimensional Geometrical surface, empty from any matter,
bounded by a closed superconducting wire, would irradiate in the
same manner as a material black-body with similar radiating area
and shape. When the superconductor that is drawing the frontier of
the geometrical surface in question, becomes normal, above $T_c$,
this radiation would disappear. This would be a direct consequence
of the quantization of geometric areas delimited by SCing
materials at the Planck-Einstein scale, and of the electromagnetic
dark energy content of the superconductor. Ultimately this would
represent different equivalent phenomenological manifestations of
the spontaneous violation of the principle of general covariance
in superconductors which, as is well known, would also lead to a
violation of energy conservation, in the covariant sense.


\begin{thebibliography}{99}


\bibitem{Rovelli} C. Rovelli, "Quantum Gravity", Cambridge
University press, (2005), pp 17-22, 246-250, 301-315


\bibitem{Tajmar01} M. Tajmar, C. J. de Matos, "Gravitomagnetic field of a rotating
superconductor and of a rotating superfluid" {\it Physica C} {\bf
385}, {551} (2003).

\bibitem{Tajmar02} M. Tajmar, C. J. de Matos, "Extended analysis of gravitomegnetic fields
in rotating superconductors and superfluids" {\it Physica C} {\bf
420}, {56} (2005).

\bibitem{de Matos01} C. J. de Matos, M. Tajmar, "Gravitomagnetic London moment and
graviton mass inside a superconductor" {\it Physica C} {\bf
432}, 167 (2005).

\bibitem{de Matos02} C. J. de Matos, "Gravitoelectromagnetism and
dark energy in superconductors", {\it Int. Mod. Phys. D}, {\bf
16}, 12B, 2599 (2007)

\bibitem{de Matos03} C. J. de Matos, "Electromagnetic dark energy and
gravitoelectrodynamics of superconductors", {\it Physica C}
{\bf 468}, 210, (2008)

\bibitem{de Matos04} C. J. de Matos, C. Beck, "Possible measurable
effects of dark energy in rotating superconductors", arXiv:
0707.1797 [gr-qc], to appear in {\it Microgravity Science and
Technology Journal}, (2008)

\bibitem{Beck01}
C. Beck, C.J. de Matos, "The dark energy scale in superconductors:
Innovative theoretical and experimental concepts", to appear in
Int J. Mod. Phys. D, arXiv:0709.2373 [gr-qc]


\bibitem{Beck02} C. Beck, M. C. Mackey, "Could dark energy be measured in the lab?" {\it Phys. Lett. B} {\bf
605}, 295 (2005).

\bibitem{Beck03} C. Beck, "Laboratory tests on dark energy" {\it J.Phys.Conf.Ser.} {\bf 31}, 123 (2006).

\bibitem{Beck04} C. Beck, M.C. Mackey, "Zeropoint fluctuations and dark energy in Josephson junctions"
, Fluctuation and Noise Letters 7, C27 (2007), astro-ph/0603067

\bibitem{Beck05} C. Beck, M.C. Mackey. "Electromagnetic dark energy", To appear in Int. J.
Mod. Phys. D 2008.

\bibitem{Beck06} C. Beck, M. C. Mackey, "Measurability of vacuum fluctuations and dark energy", {\it Physica A} {\bf 379}, 101
(2007)



\bibitem{Tate01}J. Tate, B. Cabrera, S.B. Felch, J.T. Anderson, "Precise
determination of the Cooper-pair mass", {\it Phys. Rev. Lett.}
{\bf 62} (8) 845–848 (1989).

\bibitem{Tate02} J. Tate, B. Cabrera, S.B. Felch, J.T. Anderson, "Determination
of the Cooper-pair mass in niobium", {\it Phys. Rev. B} {\bf 42}
(13) 7885–7893 (1990).

\end{thebibliography}
\end{document}